\documentclass[12pt]{elsarticle}
\usepackage{slashed}
\usepackage{epsfig,graphics}
\usepackage{hyperref}
\usepackage{bbm}
\usepackage{mathtools}
\usepackage{amssymb}
\usepackage{mathrsfs}
\usepackage{amsmath}
\usepackage{natbib}

\newlength{\llslash}

\newcommand{\tr}{\rm tr \,}

\begin{document}

\begin{frontmatter}
\title{On light vector mesons \\and chiral SU(3) extrapolations}
\author[GSI]{Xiao-Yu Guo}
\author[GSI,TUD]{M.F.M. Lutz}
\address[GSI]{GSI Helmholtzzentrum f\"ur Schwerionenforschung GmbH,\\
Planck Str. 1, 64291 Darmstadt, Germany}
\address[TUD]{Technische Universit\"at Darmstadt, D-64289 Darmstadt, Germany}
\begin{abstract}
A chiral extrapolation of the light vector meson masses in the up, down and strange quark masses of QCD is  
presented. We apply  an effective chiral Lagrangian based on the hadrogenesis conjecture to QCD lattice ensembles of PACS-CS, QCDSF-UKQCD and HSC in the strict isospin limit. The leading orders low-energy constants are determined upon a global fit to the lattice data set. We use the pion and kaon masses as well as the size of the finite volume as lattice ensemble  parameters only. The quark mass ratio on the various ensembles are then predicted in terms of our set of low-energy constants. An accurate reproduction of the vector meson masses and quark-mass ratios as provided by the lattice collaborations and the Particle Data Group (PDG) is achieved. Particular attention is paid to the $\omega-\phi$ mixing phenomenon, which is demonstrated to show a strong quark mass dependence. 
\end{abstract}
\end{frontmatter}

\section{Introduction}

What is the role of dynamical light vector mesons in low-energy QCD? In this work we wish to shed more light on this burning issue  
by a numerical application of the hadrogenesis Lagrangian to QCD lattice data on the masses of vector mesons at unphysical quark masses. 
The Lagrangian conjectures a particular role of the light vector mesons, which may be justified by a hidden scale in QCD that would arise in its
large-$N_c$ limit \cite{Lutz:2008km,Terschlusen:2012xw,Bavontaweepanya:2018yds}. 
In order to appreciate such a possibility the reader may consider chiral QCD with vanishing up, down and strange quark masses in this limit. Here the mass of the lightest meson state with $J^P \neq 0^-, 1^-$ defines a specific scale, $\Lambda_{\rm HG}$, of large-$N_c$ QCD. If that scale is large enough with $ \Lambda_{\rm HG} \sim  (2-3) $ GeV  the light vector mesons could play that particular role in low-energy QCD. Such a scale separation is hidden, since it would be masked at the physical value of $N_c=3$, but be manifest at sufficiently large values of $N_c$.

In a recent work some of the authors considered the hadrogenesis Lagrangian at the one-loop level and derived specific parameter correlations that makes the Lagrangian renormalizable order-by order in a computation of the vector meson masses \cite{Bavontaweepanya:2018yds}. This analysis relies on the large-$N_c$ assumption, 
where the chiral limit value of the vector meson masses, $M \sim N_c ^0$, is considered to be much smaller than  the chiral symmetry breaking scale, $4\pi f \sim \sqrt{N_c}$. Since at the 
physical value $N_c = 3$ such an assumption may cause poor convergence properties, a particular renormalization 
procedure was suggested that is supposed to accelerate the convergence properties by a systematic summation of terms proportional to $(M/(4\pi f))^n$. 
A further instrumental ingredient is the derivation of loop effects in terms of physical  meson masses rather than bare masses as requested by formulations of conventional 
effective field theories \cite{Lutz:2018cqo,Guo:2018kno}. It was illustrated that renormalization-scale invariant results can be obtained upon a specific recast of the counter term contributions that 
then balance the scale dependence of the loop contributions \cite{Bavontaweepanya:2018yds}. 

The quark-mass dependence of the light vector meson masses has been considered before in various works \cite{Jenkins:1995vb,Klingl:1996by,Birse:1996hd,Bijnens:1997ni,Bijnens:1997rv,Cirigliano:2003yq,Rosell:2004mn,Bruns:2013tja}. 
We consider lattice data on meson masses from PACS-CS, QCDSF-UKQCD and HSC, \cite{PACS-CS,QCDSF-UK,HSC,Dudek:2013yja,Bulava:2016mks,Brett:2018jqw}. 
There is also a data set based on the mixed action approach of LHPC \cite{LHPC}. As we have demonstrated previously for the light baryon masses  the LHPC results appear largely incompatible  with corresponding results from further lattice groups, in particular from PACS-CS\cite{Lutz:2018cqo}. We confirm a similar pattern also in the vector meson sector and therefore do not consider the LHPC data set any further. 

\section{Chiral dynamics from the hadrogenesis Lagrangian}

In the following we briefly recall the relevant terms of the hadrogensis Lagrangian \cite{Lutz:2008km,Terschlusen:2012xw,Bavontaweepanya:2018yds}. It is formulated in terms of the anti-symmetric 
tensor field $\Phi_{\mu \nu}$ that interpolate the light vector mesons together with the chiral building blocks $U_\mu$ and $\chi_\pm$ that contain the fields of the Goldstone bosons $\Phi$ and the quark mass parameters of QCD.  
The covariant derivative is introduced with $D_\mu$. We have 
\allowdisplaybreaks
\begin{eqnarray}
&& {\mathcal L}^{(3)}_2  =  \frac{i}{2}\,f\, h_1\,{\rm tr}\,\Big\{U_\mu\,\Phi^{\mu\nu}\,U_\nu\Big\}
   +\ \frac{i}{8} \,h_2 \,\varepsilon^{\mu\nu\alpha\beta}\,
  {\rm tr} \,\Big\{ \big[\Phi_{\mu \nu},\,(D^\tau \Phi_{\tau \alpha})\big]_+ \,U_\beta\Big\}
\nonumber\\
&&\qquad   -\, \frac{i}{4} \,\frac{M^2}{f} \, h_3 \,{\rm tr}\,\Big\{
  \Phi_{\mu \tau}\,\Phi^{\mu \nu}\,\Phi^{\tau}_{\;\;\, \nu} \Big\}\,,
\nonumber \\ \nonumber\\
&& {\mathcal L}^{(4)}_2  = 
    \frac{1}{8} \,
    {\tr } \,\Big\{  g_1 \,\big[ \Phi_{\mu \nu }\,,U_\alpha \big]_+ \, \big[U^\alpha, \Phi^{\mu \nu} \big]_+
    +  g_2 \,
   \big[ \Phi_{\mu \nu }\,,U_\alpha \big]_- \, \big[U^\alpha, \Phi^{\mu \nu}\big]_- \Big\}
\nonumber \\
&& \qquad
    + \,\frac{1}{8} \, 
    {\tr } \,\Big\{g_3 \,\big[\,U_\mu\,,U^\nu \big]_+ \, \big[\Phi_{\nu \tau}\,, \Phi^{\mu \tau}\big]_+  
       +g_4 \,\big[\,U_\mu\,,U^\nu \big]_- \, \big[\Phi_{\nu \tau} \,, \Phi^{\mu \tau}\big]_- \Big\}
\nonumber \\
&& \qquad +\, \frac{1}{8} \, g_5 \,
    {\tr } \,\Big\{\big[\Phi^{\mu \tau} , U_\mu\big]_- \, \big[\Phi_{\nu \tau} \,, U^\nu\big]_- \Big\}
    \,,
\nonumber\\    
&& \mathcal{L}_{4}^{(V)} =  
\frac{e_2}{8}\,M^4\,{\tr\{\Phi_{\mu\nu}\}\,\tr\{\Phi^{\mu\nu}\} }
+ \frac{b_1}{8}\, M^2\,{\tr } \,\Big\{ \Phi^{\mu \nu}\,\Phi_{\mu \nu}\,\chi_+\Big\} 
\nonumber\\
&& \qquad + \, \frac{b_2}{8}\,M^2 \,{\tr\{\Phi_{\mu\nu}\,\Phi^{\mu\nu}\}\,\tr\{\chi_+\} }
+\,  \frac{b_3}{8}\,M^2\,{\tr\{\Phi_{\mu\nu}\}\,\tr\{\Phi^{\mu\nu}\,\chi_+\}}
\nonumber\\ 
&& \qquad + \, \frac{c_1}{8}\,{\tr\,\{\Phi_{\mu\nu}\,\chi_+\,\Phi^{\mu \nu}\,\chi_+\}} +  \frac{c_2}{8}\,\tr\,\{\Phi_{\mu\nu}\,\Phi^{\mu\nu}\,\chi_+^2\} 
\nonumber\\ 
&& \qquad + \,
\frac{c_3}{8}\,{\tr\,\{\Phi_{\mu\nu}\,\Phi^{\mu\nu}\}\,\tr\,\{\chi_+^2\}} + \frac{c_4}{8}\,\tr\,\{\Phi_{\mu\nu}\,\Phi^{\mu\nu}\,\chi_+\}\,\tr\,\{\chi_+\} 
\nonumber\\
&& \qquad  + \,   \frac{c_5}{8}\,\tr\,\{\Phi^{\mu\nu}\,\chi_+\}\,\tr\,\{\Phi_{\mu\nu}\chi_+\} +\frac{c_6}{8}\,\tr\,\{\Phi^{\mu\nu}\}\,\tr\,\{\Phi_{\mu\nu}\,\chi_+^2\}
\,,    
\label{def-L24}
\end{eqnarray}
where we recall 
\begin{eqnarray}
&& U_\mu  =  {\textstyle \frac{1}{2}}\,e^{-i\,\frac{\Phi}{2\,f}} \left(
    \partial_\mu \,e^{i\,\frac{\Phi}{f}} \right) e^{-i\,\frac{\Phi}{2\,f}} \,, \qquad  \quad
    D_\mu \Phi_{\alpha\beta}  =  \partial_\mu \Phi_{\alpha\beta} + \big[\Gamma_\mu,\,\Phi_{\alpha\beta}\big]_- \,,
\nonumber\\
&& \Gamma_\mu  = {\textstyle \frac{1}{2}}\,e^{-i\,\frac{\Phi}{2\,f}} \,
\partial_\mu  \,e^{+i\,\frac{\Phi}{2\,f}}
+{\textstyle \frac{1}{2}}\, e^{+i\,\frac{\Phi}{2\,f}} \,
\partial_\mu \,e^{-i\,\frac{\Phi}{2\,f}}\,,    
\nonumber\\
&& \chi_\pm = {\textstyle \frac{1}{2}} \,
e^{+i\,\frac{\Phi}{2\,f}} \,\chi_0 \,e^{+i\,\frac{\Phi}{2\,f}}
\pm {\textstyle \frac{1}{2}}\,  e^{-i\,\frac{\Phi}{2\,f}} \,\chi_0 \,e^{-i\,\frac{\Phi}{2\,f}} \,,
\label{def-chi}
\end{eqnarray}
with $\chi_0 =2\,B_0\, {\rm diag} (m_u,m_d,m_s)$.  Note that as in \cite{Bavontaweepanya:2018yds} we do not yet consider the explicit effects of the $ \eta'$ field \cite{Terschlusen:2012xw}.  

The order of a given interaction term is determined in two steps. First we factor out a term $1/f^n \sim 1/\sqrt{N_c}^{n}$, where $n$ is fixed such that the rescaled vertex acquires its expected large-$N_c$ scaling behaviour from QCD. The residual coupling constant $g$ carries some dimension $d$. While for the case $d < 0$ we expect $g \sim \Lambda_{\rm HG}^d$ the case $d> 0$ implies $g \sim M^d $ with $M$ beeing the chiral limit mass of the light vector meson masses. 

A lower bound for the power-counting order of the vertex is then implied by the formal counting rule $M \sim D_\mu  \sim U_\mu \sim Q $ and $\chi_\pm \sim Q^2$. The renormalization condition may yet imply an increase for the formal power of the considered  vertex  \cite{Bavontaweepanya:2018yds}. For instance this was shown to be unavoidable for the symmetry breaking terms $b_n$ in (\ref{def-L24}). The latter can be considered consistently only if introduced at order $Q^4$ rather than $Q^2$ as one may expect naively. Similarly it is necessary to impose the 
two sum rules
\begin{eqnarray}
&&g_3 = \frac{1}{4}\,h_2^2 -4\,g_1\,, \qquad \qquad \qquad  g_5 = g_3 +4\,g_2\,,  
\label{sum-rules-gi}
\end{eqnarray}
at leading order in the power counting scheme. The parameter $g_4$ does not enter the vector meson masses at the one-loop order, however, it is anticipated that an extended one-loop analysis leads to the additional relation $4\,g_4 = g_5 - h_2^2/4$.

Once we insist on the  relations (\ref{sum-rules-gi}) the  parameters $c_{1-6}$  remain renormalization scale invariant. Moreover, the effect of the parameters $g_1$ and $g_2$ on the vector meson masses vanishes in the infinite volume limit identically. 
We note also that in (\ref{def-L24}) we consider all $c_n$ terms to be relevant at $Q^4$ despite the fact that the terms proportional to $c_{3,4,5,6}$ are suppressed by a factor $1/N_c$ as compared to $c_{1,2}$ due to the presence of double flavour traces. Without that assumption 
it is not possible to arrive at a reproduction of the vector meson masses as provided by the various lattice groups.

Some of the low-energy parameters have been estimated before in \cite{Lutz:2008km,Terschlusen:2012xw,Bavontaweepanya:2018yds} with
\begin{eqnarray}
&&h_1 = \frac{2.5 \pm 0.25}{90\,{\rm MeV}} \,f\,\,, \qquad \qquad h_2 =  \frac{2.33\pm 0.03}{ 90\,{\rm MeV}\,\cos \epsilon_{\omega }}\,f\,,\qquad 
\nonumber\\
&& h_3 =  \frac{0.05}{90\,{\rm MeV}} \,f \,,
\label{def-hi}
\end{eqnarray}
in terms of the chiral limit value of the pion and kaon decay constants $f$. The estimate for $h_2$ involves the $\phi -\omega $ mixing angle, $\epsilon_\omega$, evaluated at the on-shell $\omega$ meson mass 
as explained in \cite{Bavontaweepanya:2018yds}. For the remaining low-energy constants no significant estimates are available so far.

Given the effective Lagrangian (\ref{def-L24}) it is straight forward to derive the contributions to the Goldstone boson and vector meson polarization tensors $\Pi_H=\Pi_H (s= M_H^2)$, where $H$ stands for either 
a Goldstone boson $H = P$ or a vector meson $H = V$. We consider here all terms up to order $Q^4$ in our counting scheme. For a discussion of such a computation in a finite box we refer to \cite{Bavontaweepanya:2018yds}, where 
the various 'tree-level', 'tadpole' and 'bubble' type contributions are specified in terms of the low-energy constants as recalled in (\ref{def-L24}).  This gives rise to a set of coupled and non-linear mass equations of the following form
\begin{eqnarray}
&& M^2_H -  \Pi_H^{\rm tree-level}  - \Pi^{\rm tadpole}_H - \Pi^{\rm bubble}_H/ Z_H  = 0\,, \qquad  
\label{gap-equation}
\end{eqnarray}
where we consider the wave-function factor $Z_H$, in the specific form as suggested in \cite{Lutz:2018cqo,Guo:2018kno}
\begin{eqnarray}
&& Z_H = \Big( 1 +\frac{\partial }{\partial s}\,\Pi^{\rm bubble}_H  \Big)/  \Big( 1 -\frac{\partial }{\partial s}\,\Pi^{\rm tree-level}_H  \Big)\,.
\label{def-ZH}
\end{eqnarray}
With (\ref{def-ZH}) the residuum of the propagator pole is normalized to one always.

While the form of the tree-level contributions is given in Tab. III and Eq. 51 of \cite{Bavontaweepanya:2018yds}, 
the tadpole contributions are detailed in Tab. I and Eq. 16 for the vector mesons, and in Tab. IV and Eq. 39 for the Goldstone bosons\footnote{Given the form of the bubble loop function in (\ref{bubble-V})  the renormalized parameters 
$g^r_1 = g_1 +h_2^2/8$, $g^r_2 =g_2$, $g^r_3 = g_3 -3\,h_2^2/4$ and $g^r_5 = g_5 -3\, h_2^2/4$ have to be taken in Tab. I of \cite{Bavontaweepanya:2018yds}.  }.  The scale dependent tadpole integrals,  $\bar I_Q$ and  $\bar I^{(2)}_Q$  with $Q\in [8]$ running over the Goldstone bosons take 
into account finite volume effects. They can be taken from the previous work \cite{Lutz:2014oxa}. We recall that in the infinite volume limit its holds $\bar I^{(2)}_Q \to m_Q^2\,\bar I_Q/4$. 

It is left to discuss the vector meson tadpole contributions and the bubble loop contributions, which owing to the Passarino-Veltman reduction, can be expressed in terms of tadpole integrals and the scalar 
bubble functions $I_{AB}$ with  $A,B \in \{P, V\}$. As proven in  \cite{Lutz:2014oxa} this 
reduction scheme can be generalized to the finite box case. Following the strategy proposed in \cite{Lutz:2018cqo,Guo:2018kno} the scalar bubble function turns regulator scale invariant after a proper renormalization. The key observation is that given the particular renormalization scheme all tadpole integrals involving a vector meson  mass can be dropped.
This is expected to be justified at least in the chiral domain with $m_{\pi, K, \eta} \ll M$. Here the vector mesons should be considered as heavy fields, which are known to generate power-counting violating contributions in that case. In particular any vector meson tadpole contribution must be dropped as to arrive at a strict realization of counting rules in dimensional regularization. This amounts to a particular renormalization procedure, where it is guaranteed by the  Passarino-Veltman reduction that it is in compliance with the chiral Ward identities of QCD.

Given the fact that it is justified to drop contributions from any vector meson tadpole, we can remove the renormalization scale dependence of the scalar bubble functions $I_{AB}$ with $A,B \in P, V$ by subtraction terms 
involving such vector meson tadpoles \cite{Bavontaweepanya:2018yds}. Here we provide the final expressions used in the numerical application of our work. For the pseudo-scalar (P) and vector mesons (V) we apply 
the generic expressions
\begin{eqnarray}
&& \Pi^{\rm bubble}_{P\in [8]}  =  \sum_{\substack{Q\in [8],\,V\in [9]}}
\Big(\frac{G^{(P)}_{QV}}{4\,f}\Big)^2 \Bigg\{ -\Big(M_V^2-m_Q^2\Big) ^2\,\Delta I_{QV} 
-m_P^2 \,\hat I^V_Q \nonumber\\
&& \qquad \qquad  \qquad \qquad \qquad \qquad \quad \!
-\,m_P^2\,\Big( m_P^2 -2\,(m_Q^2+M_V^2)\Big)\,\bar I_{QV}\Bigg\}
\nonumber\\
&& \qquad \quad +\,  \sum_{V,R \,\in [9]}
\Big(\frac{G_{VR}^{(P)}}{4\,f}\Big)^2\Bigg\{  -  \alpha^{P}_{VR}\,\Big( M_R^2-M_V^2\Big)^2\,\Delta I_{VR}
\nonumber\\
&& \qquad  \qquad \qquad  \qquad \qquad \quad \;\;- \,\alpha^{P}_{VR}\,m_P^2\,\Big( m_P^2-2\,(M_R^2+M_V^2)\Big) \,\bar I_{VR} \Bigg\}\,,
\nonumber\\
&&\alpha_{VR}^{P}=  \frac{(M_V^2+M_R^2 )^2 }{4\,M_V^2\, M_R^2}  \,,  
\label{bubble-P}
\\ \nonumber \\
&& \Pi^{\rm bubble}_{V\in [9]}  =  \sum_{Q,P\in [8]} 
\Bigg(\frac{G_{QP}^{(V)}}{4\,f}\Bigg)^2 \Bigg\{ -\Big(m_P^2-m_Q^2\Big)^2\,\Delta I_{QP} -M_V^2\,\Big( \hat I^V_Q + \hat I^V_P\Big)
\nonumber\\
&& \qquad \qquad \qquad   \quad -\, M_V^2\,\Big( M_V^2 -2\, (m_P^2+ m_Q^2)\Big)\,\bar I_{QP} \Bigg\}
\nonumber\\
&& \qquad \quad + \sum_{Q\in [8] ,\,R\in [9]} \,\Bigg(\frac{G_{QR}^{(V)} }{4\,f}\Bigg)^2
\Bigg\{ - \alpha^{V}_{QR}\,
\Big( M_R^2-m_Q^2\Big)^2\,\Delta I_{QR} 
 \nonumber\\
&& \qquad \qquad \qquad   \quad  -\, M^2_V  \, \hat I^R_Q  
- \alpha^{V}_{QR}\,
M_V^2\,\Big( M_V^2- 2\,(m_Q^2+M_R^2)\Big)\,\bar I_{QR} \Bigg\} 
\nonumber\\
&& \qquad \quad  + \sum_{R,T\in [9]} \Bigg(\frac{G_{R \,T}^{(V)}}{4\,f}\Bigg)^2
\Bigg\{ - \alpha_{R\,T}^{V}\,\Big( M_R^2-M_T^2\Big)^2 \Delta I_{RT}
\nonumber\\
&& \qquad \qquad \qquad \quad -\, \alpha_{R\,T}^{V}\,M_V^2\Big( M_V^2-2\,(M_R^2+M_T^2)\Big)\,\bar I_{R\,T}\Bigg\}\,,
\nonumber\\ \nonumber\\
&& \alpha^{V}_{QR} = \frac{ (M_R^2+M_V^2)^2 }{4\, M_R^2\,M_V^2}  \,, \qquad \qquad \qquad  \alpha^{V}_{R\,T} = M^4\,\frac{M_R^2+M_V^2+M_T^2}{3\,M_R^2\,M_V^2\,M_T^2}\,,
\label{bubble-V}
\end{eqnarray}
in terms of the scalar bubble functions  $ \Delta I_{AB} $, $\bar I_{AB}$ together with a renormalized tadpole integrals $\hat I^R_Q$ that depends on the pseudo-scalar meson mass $m_Q$ and the vector meson mass $M_R$. 
The detailed form of the subtractions in the various terms is provided
in \cite{Lutz:2018cqo,Guo:2018kno}. The finite volume parts can be taken from \cite{Lutz:2014oxa}. We note that, while by construction it holds $\Delta I_{AB}(s = 0) = 0$ for any A, B, the 
subtraction in $\bar I_{AB}$ depends on the type of A and B. 

The Clebsch coefficients $G_{AB}^{(H)}$ in (\ref{bubble-P}) and (\ref{bubble-V}) as given in Tab. II and Tab. V of \cite{Bavontaweepanya:2018yds} depend on the coupling constants $h_{1-3}$ only. They are 
computed with respect to  bare $\omega$ and $\phi$ meson states with either only up and down 
or only strange quark content. The physical $\omega$ and $\phi$ meson may acquire a more complicated flavour structure. This is a direct consequence of the non-vanishing of the transition polarization 
tensor $\Pi_{\omega \phi}(s)\neq 0$ \cite{Bruns:2013tja,Bavontaweepanya:2018yds}. 
One may introduce an $\omega-\phi$ mixing angle $\epsilon$ by 
\begin{eqnarray}
&& \omega =\omega' \, \cos \epsilon +\phi' \,\sin \epsilon \,, \qquad \qquad \qquad \phi =\phi'\,  \cos \epsilon  -\omega'\, \sin \epsilon\,,
\nonumber\\
&& \qquad{\rm with} \qquad \Pi_{\omega \phi} =\frac{1}{2}\,\Big(\Pi_\phi - \Pi_\omega \Big)\,\tan (2\,\epsilon) \,,
\label{def-mixing}
\end{eqnarray}
where  the transformed fields $\omega'$ and $\phi'$ are the physical fields related to the mass eigenstates.

There are different strategies on how to determine 
the mixing angles. In the $\omega- \phi$ basis the determination of the $\omega $ and $\phi$ meson masses requires the consideration of a two dimensional polarization tensor with the off-diagonal elements 
given by $\Pi_{\omega \phi}(s)$. Once the physical $\omega $ and $\phi $ meson masses are determined one may infer from (\ref{def-mixing}) two mixing angles $\epsilon_\omega$ and $\epsilon_\phi$, evaluated 
at either $s = M^2_\omega$ or $s = M_\phi^2$ respectively. 

Equivalent to this procedure is that we introduce the two mixing angles from the beginning and determine their values by the request that 
$\Pi_{\omega' \phi'}(s)$ vanishes at both the physical $\omega$ meson but also at the physical $\phi$ meson mass. With this ansatz the two dimensional matrix structure is factorized into its $\omega'$ and $\phi'$ 
components. In our previous work and here we choose the latter approach. Given our one-loop level 
we argue that then it is justified to approximate the vector meson propagators used inside the one-loop integral by a simple pole term with a pole mass set to its physical value.  

We note that there are contributions to $\Pi_{\omega' \phi'}(s)$ from tree-level, tadpole as well as from the bubble term.  
The latter contribution follows from $\Pi^{\rm bubble}_V $ upon the replacement 
\begin{eqnarray}
\big(G_{QP}^{(V)}\big)^2\to G_{QP}^{(\omega)}\,G_{QP}^{(\phi)}\qquad  {\rm and} \qquad  M_V^2 \to s \,,
\label{replace-bubble}
\end{eqnarray}
in all three sums of (8).

\section{Global fit to meson masses from QCD lattice simulations }

We discuss our strategy how to use the QCD lattice data following our previous works \cite{Lutz:2014oxa,Lutz:2018cqo,Guo:2018kno}. A subset of low-energy constants is 
fixed by the requirement that  the isospin averaged vector meson masses are reproduced as provided by the PDG \cite{Patrignani:2016xqp}. In addition we insist on reproducing the empirical $\omega-\phi$ mixing angle at the 
$\phi$ mass, with $|\epsilon_\phi | \simeq 3.32^{\,\circ}$ as  derived from the decay $\phi \to \pi_0\, \gamma$ in \cite{Klingl:1996by}. With this we determine the 5 parameters $b_1, b_2, b_3, c_1, e_2$. 
The products $B_0\,m, B_0 \,m_s$ and $L_8 + 3\,L_7$ are set by the request to reproduce 
the empirical pion, kaon and eta masses. That leaves 11 free parmeters only to be adjusted to the lattice data. They are $f, M, c_1, c_3, c_4, c_5, c_6, g_1, g_2$ and $L_4 - 2\,L_6, L_5 -2\,L_8$.
To actually perform the fits is a computational challenge. For any set of the low-energy parameters nine coupled non-linear equations are to be solved on each lattice ensemble  
considered. We apply the evolutionary algorithm of GENEVA 1.9.0-GSI \cite{Geneva} with runs of a population size 1500 on 300 parallel CPU cores.

In our global fit to the QCD lattice data on vector meson masses we consider results from PACS-CS, QCDSF-UKQCD and HSC \cite{PACS-CS,QCDSF-UK,HSC,Dudek:2013yja,Bulava:2016mks,Brett:2018jqw}. To be more precise we use the energy levels as measured on various QCD lattice ensembles in a finite box. This is justified since in our computations we incorporate the finite volume effects systematically. For the considered ensembles the volumes are quite small so that there is typically only one energy level that is relevant. Since we do not consider discretization effects and also have a residual uncertainty in our one-loop chiral extrapolation approach we assign each such energy level a systematical error that is added to its statistical error in quadrature. Our ansatz for the systematical error is asymmetric since the asymptotic determination of an energy level from a correlation function measured on a considered lattice ensemble may sometimes provide an upper estimate for its energy level only. This is so if the statistical error gets large before the true asymptotic exponential tail is reached. As has been shown by HSC this is particularly troublesome for the $\rho$ and $K^*$ for which accurate results may not be reached in terms of interpolating operators with quark and antiquark fields only. Here the additional source functions with four quark field operators may be required.  Our chisquare function, $\chi^2$, assumes a universal but asymmetric systematical error for the vector meson energy levels, where the lower error is chosen twice as large as the upper error. Its size will be chosen to arrive at about $\chi^2/N \sim 1$, with $N$ the number of fitted vector meson masses.

\begin{table}[t]
\setlength{\tabcolsep}{3.2mm}
\renewcommand{\arraystretch}{1.15}
\begin{center}
\begin{tabular}{l||rrr||r} 
                             & Fit 1       & Fit 2    & Fit 3 & Literature   \\ \hline \hline
$f$ [MeV]                    &  73.57      &    70.72   &   67.51  &  64 - 71 \cite{Bijnens:2014lea} \\
$M$ [MeV]                    &  759.3      &    758.8   &   757.0  &  618 - 696 \cite{Bruns:2013tja} \\ \hline 
$e_2$ [GeV$^{-2}$]           & -0.1072     &  -0.0890   & -0.0930  &   \\
$b_1$ [GeV$^{-2}$]           &  1.2224     &   1.3420   &  1.4009   &  \\
$b_2$ [GeV$^{-2}$]           & -0.2042     &  -0.1686   & -0.1868  &  \\
$b_3$ [GeV$^{-2}$]           &  0.5131     &   0.3469   &  0.4151  &  \\\hline       
$g_1$                        & -0.1180     &   0.4038   &  0.5025  &  \\
$g_2$                        & -0.9657     &  -0.3173   & -0.3840  &   \\ \hline
$c_1$ [GeV$^{-2}$]           & -0.3642     &  -0.3876   & -0.3771  &   \\ 
$c_2$ [GeV$^{-2}$]           & -1.0143     &  -1.1910   & -1.1962  &   \\ 
$c_3$ [GeV$^{-2}$]           &  0.5557     &   0.5366   &  0.5513 &   \\ 
$c_4$ [GeV$^{-2}$]           &  0.3803     &   0.3833   &  0.3449  &   \\ 
$c_5$ [GeV$^{-2}$]           & -0.1267     &  -0.1325   & -0.1525  &   \\ 
$c_6$ [GeV$^{-2}$]           &  0.0221     &   0.1656   &  0.0853  &   \\ \hline
$(L_4-2\,L_6)\times 10^3$    &  0.0977     &  0.0119    &  0.0469  &  -0.22  - \, 0.02 \cite{Bijnens:2014lea}\\
$(L_5 -2\,L_8)\times 10^3$   &  0.5171     &  0.5209    &  0.4144  &  \!\! 0.07  -\,  0.16  \cite{Bijnens:2014lea} \\
$(L_8 +3\,L_7)\times 10^3$   & -0.4119     & -0.3912    & -0.3416  &  -0.55 - -0.40 \cite{Bijnens:2014lea}  \\ \hline \hline
$a_{\rm PACS-CS}$ [fm]       & 0.0914      &  0.0924    & 0.0919 &  0.0907(13) \cite{PACS-CS} \\
$\chi^2 /N$                  & 0.26        &  0.70      & 0.54   &                             \\ \hline
$a_{\rm QCDSF-UKQCD}$ [fm]   & 0.0730      &  0.0717    & 0.0722 &   0.0765(15) \cite{QCDSF-UK} \\
$\chi^2 /N$ : $N_L = 32 $    & 0.61        &  0.97     & 0.84   &   0.0740(4)\phantom{0} \cite{Bornyakov:2016dzn}  \\ 
$\chi^2 /N$ : $N_L = 24 $    & 0.64        &  0.80      & 0.50   &    \\ 
$\chi^2 /N$ : $N_L = 16 $    & 0.01        &  0.05      & 0.01   &    \\ \hline
$a_{\rm HSC}$  [fm]          & 0.1202      &  0.1194   & 0.1200 &  0.1227(8)\phantom{0} \cite{HSC} \\
$\chi^2 /N$                  & 0.81        &  1.17      & 1.46   &   \\
\end{tabular}
\caption{Low-energy parameters as implied by Fit 1, Fit 2 and Fit 3 as explained in the text. At the physical point we have the $ \epsilon_\phi = 3.32^{\,\circ} $  but  $ \epsilon_\omega = \{21.9^{\,\circ}, 20.1^{\,\circ}, 19.9^{\,\circ}\} $ for the three fit scenarios respectively. The 'physical' quark masses are implied with  $2\,B_0\,m = \{1.141,1.077,1.106 \}\,m_\pi^2$ and $B_0 \,(m+ m_s) = \{1.201,1.167,1.181\} \,m_K^2$ in terms of the physical pion and kaon masses. We use $\mu =0.77 $ GeV for the renormalization scale. }
\end{center}
\label{tab:parameter}
\end{table}

\begin{figure}[t]
\centering
\includegraphics[width=0.99\textwidth]{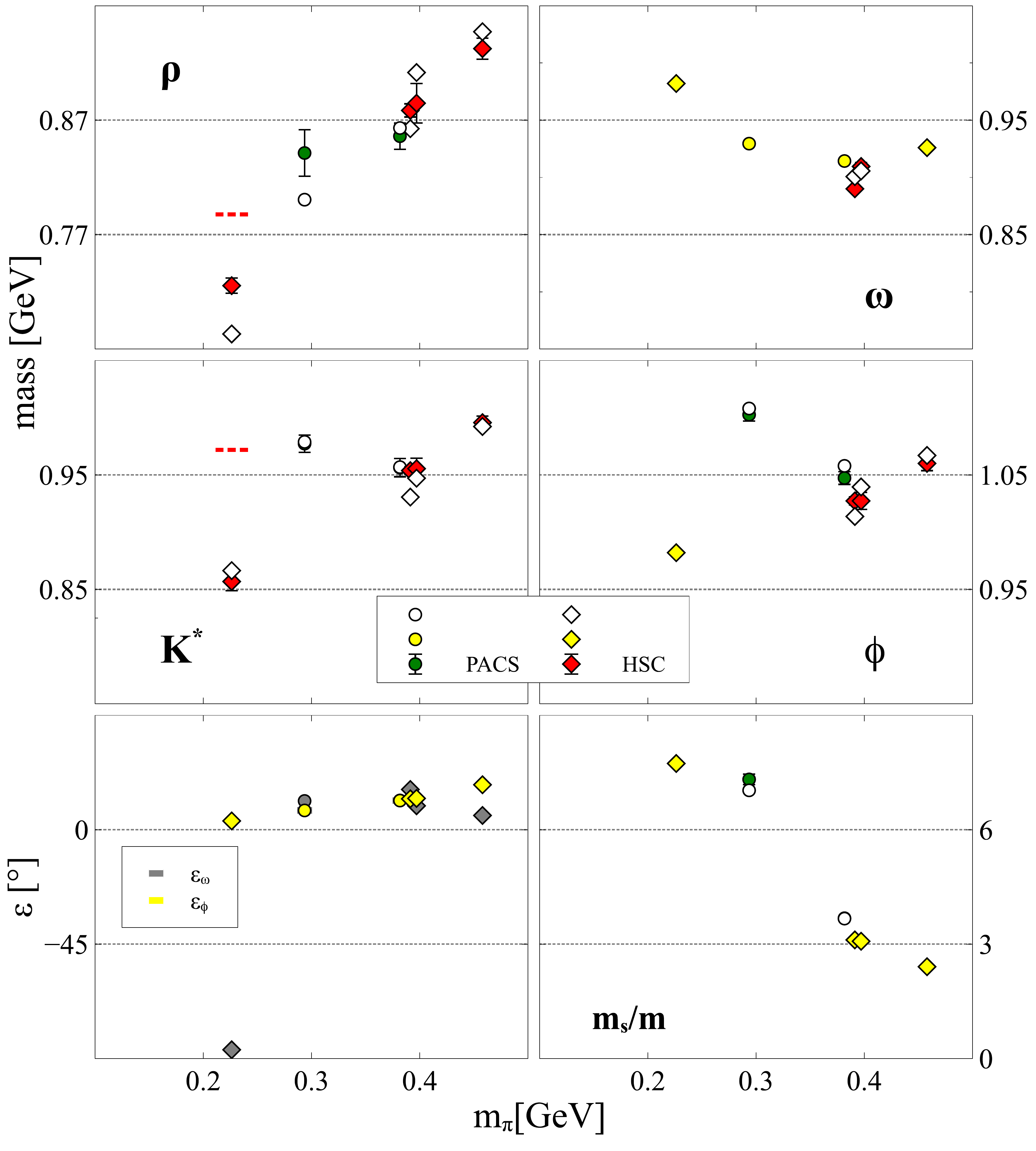}
\caption{Our results for the vector meson masses, the $\omega-\phi$ mixing angles and quark mass ratios on the PACS-CS and HSC ensembles. The lattice results are given by green (PACS-CS) and red (HSC) filled symbols, where statistical errors are shown only \cite{PACS-CS,HSC,Dudek:2013yja,Bulava:2016mks,Brett:2018jqw}. They are compared to the chiral extrapolation results in open symbols, which are always displayed on top of the lattice symbols. We use yellow or grey colour filled symbols for the cases where there is no corresponding lattice point available yet.  }
\label{fig:1}
\end{figure}

\begin{figure}[t]
\centering
\includegraphics[width=0.99\textwidth]{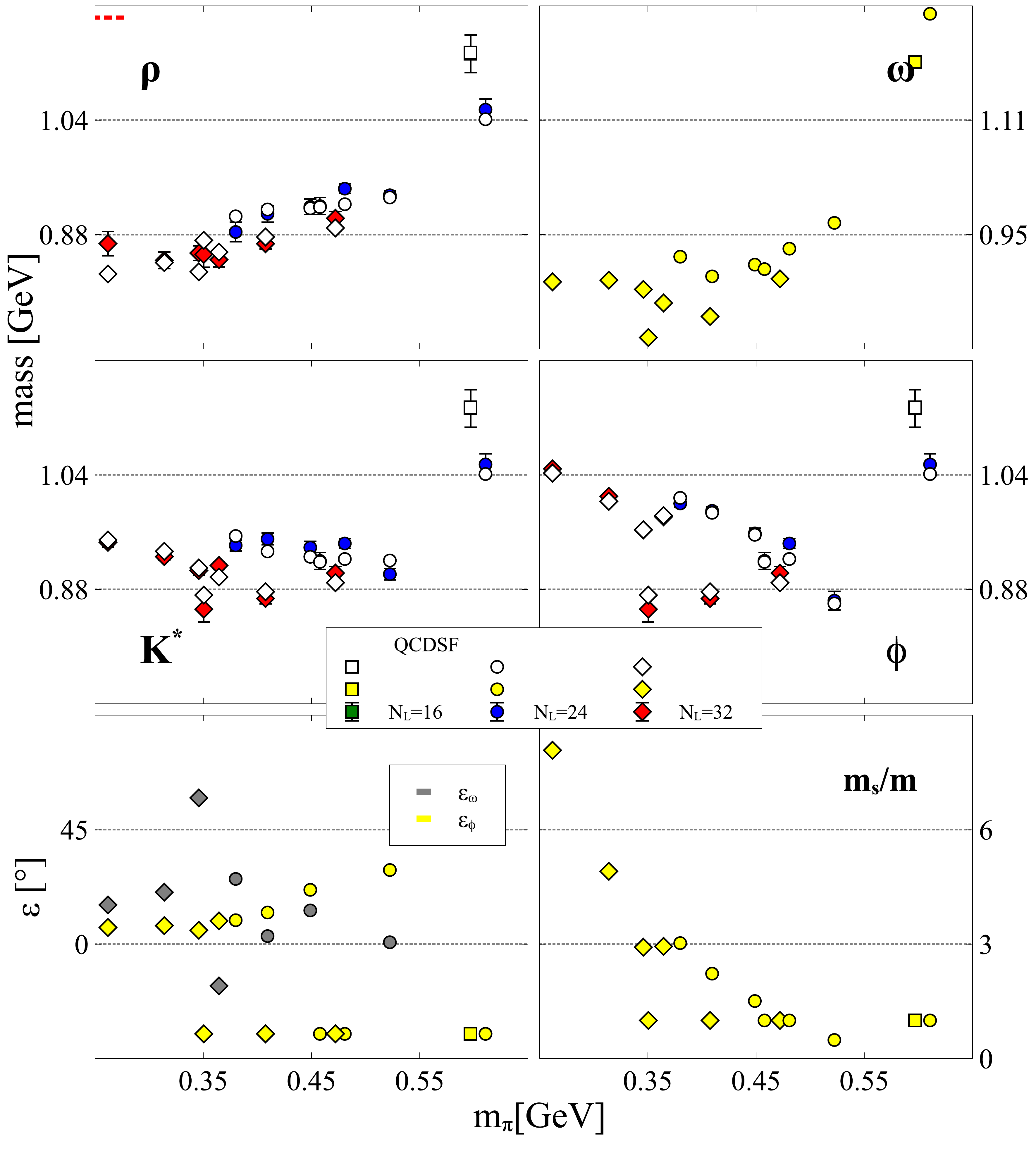}
\caption{Our results for the vector meson masses, the $\omega-\phi$ mixing angles and quark mass ratios on the QCDSF-UKQCD ensembles.  The lattice results are given by green ($16^3$ lattice), blue ($24^3$ lattice) red ($32^3$ lattice) filled symbols, where statistical errors are shown only \cite{QCDSF-UK}. They are compared to the chiral extrapolation results in open symbols, which are always displayed on top of the lattice symbols. We use yellow or grey colour filled symbols for the cases where there is no corresponding lattice point available yet. }
\label{fig:2}
\end{figure}

In Tab. 1 we collect the values of the low-energy constants according to three scenarios. For each one we set the lattice scales as required by the considered lattice collaborations. We use an ensemble of observable quantities to do so. The empirical vector meson masses from the PDG are reproduced identically in any of our global fits. While in Fit 1 we consider the vector meson masses only, in Fit 2 and Fit 3 constraints from additional lattice data on the pion and kaon decay constants, as provided by HPQCD and CLS  \cite{Dowdall:2013rya,Bruno:2016plf}, are imposed. This will be detailed in a separate work \cite{Guo:2018c}. Note that for the ensembles of HPQCD and CLS  there are rather few values for the vector meson masses available so far \cite{Andersen:2018mau}. We will return to the meson masses on those ensembles in \cite{Guo:2018c}.
The chisquare values collect in Tab. 1 are computed with respect to an 
asymmetric systematical error of 10 MeV for its upper value in the $\rho$ and $K^*$ masses as explained above. From the values in the last eight rows of Tab. 1 we conclude that such additional constraints deteriorate our description of the vector meson masses only slightly. Most of the low-energy constants are of quite similar size in the three scenarios. Exceptions are $g_1$ and $g_2$. Each of the parameter sets provides a good description of the vector meson masses. With few exceptions our LEC are compatible with estimates from previous studies. 
Our results for the spatial lattice scales in Tab. 1 are quite compatible with previous 
studies on the PACS-CS and HSC ensembles. However, this is not so on the QCDSF-UKQCD ensembles. Our values in Tab. 1 are always significantly smaller than previous estimates by QCDSF-UKQCD based mainly on their baryon masses \cite{QCDSF-UK,Lutz:2018cqo}. It would be important to trace the source of this discrepancy.

For any ensemble we take its published pion and kaon masses as input parameters. The set of nine coupled and non-linear mass equations is solved in terms of the two quark masses, $B_0\, m$, $B_0\, m_s$,  the remaining 
5 meson masses, $m_\eta, M_\rho, M_\omega, M_{K^*}, M_\phi$ and the two mixing angles $\epsilon_\omega$ and  $\epsilon_\phi $. In all our fits we consider the vector meson masses on PACS-CS, QCDSF-UKQCD and HSC  ensembles 
that correspond to $m_\pi $ and $ m_K$ smaller than 600 MeV only. 
In Fig. \ref{fig:1} and Fig. \ref{fig:2}  we show results only for our first scenario in  Tab. 1. Let us begin with Fig. \ref{fig:1} where we show the implications of the PACS-CS and HSC ensembles \cite{PACS-CS,HSC,Dudek:2013yja,Bulava:2016mks,Brett:2018jqw}. A fair reproduction of all vector meson energy levels is achieved. The $\rho$ energy levels  on the ensembles with the two lightest pion masses are not considered in our chisquare function. In both cases the first unperturbed scattering level at $E = 2\,\sqrt{m_\pi^2 + (2\,\pi/L)^2} $ (see  dashed lines in the figure) appear well separated from the driving level and therefore we expect our self consistent approach to be applicable.

We find interesting the rich pattern predicted for the two $\omega-\phi$ mixing angles. A striking dependence on the quark masses is obtained. This is not surprising since in the flavour SU(3) limit both mixing angles must be degenerate and take the value $\epsilon_\omega = \epsilon_\phi \simeq -35.26^{\,\circ}$, such that the $ \omega$ turns into a flavour singlet state. The figures show also our results for the quark mass ratio $2\,m_s/(m_u + m_d)$, which did not enter our chisquare function. Such ratios are not available from HSC, since given their asymmetric lattice set up, with distinct spatial and temporal lattice scales, it would be quite a challenge to derive values for the latter. In contrast, for the two ensembles of PACS-CS such ratios are available and they do match our predictions quite accurately.

We turn to Fig. \ref{fig:2} where ensembles of the QCDSF-UKQCD   \cite{QCDSF-UK} are scrutinized. Lattice data are available for three distinct volumes taken on $16^3$, $24^3$ and $32^3$ lattices. The energy levels of the $\rho, K^*$ and $\phi$ are well reproduced on all three volumes. Again the two $ \omega-\phi $ mixing angles are predicted to receive a striking quark-mass dependence. 
Quark-mass ratios on those ensembles are not available. We show our predictions nevertheless.

\section{Summary and outlook}

In this work we scrutinzed the hadrogenesis conjecture against QCD lattice data on the light vector meson masses.  
Based on this conjecture a chiral Lagrangian with flavour SU(3) vector meson fields was constructed recently that is expected to describe hadronic physics in the meson sector of QCD below the hadrogensis scale with $\Lambda_{HG}> 2$ GeV. We considered the strict isospin limit. A successful reproduction of the meson masses on the PACS-CS, QCDSF-UKQCDand HSC ensembles was achieved at the one-loop level based on uniform sets of low-energy constants. A striking prediction of our analysis are the $\omega-\phi$ mixing angles and the quark-mass ratios on all considered lattice ensembles. We observe that the mixing angle takes quite different values on the $\omega$ and $\phi$ mass poles. Our results for the quark-mass ratios match the ones claimed by PACS-CS on their ensembles. For QCDSF-UK and HSC such ratios were not available before. 

We will report next on an application of our framework to the pion and kaon decay constants as measured by HPQCD, CLS and ETMC on various QCD ensembles. This brings in only two additional low-energy constants, but a wealth of additional lattice data points. Also, we plan to incorporate the $\eta'$ meson into 
our approach.

\vskip0.3cm
\centerline{\bf{Acknowledgments}}
\vskip0.3cm
John Bulava and Sinead Ryan are acknowledged for stimulating discussions. M.F.M. Lutz thanks Kilian Schwarz and Jan Knedlik for support on distributed computing issues. 

\bibliographystyle{elsarticle-num}
\bibliography{literature}

\begin{thebibliography}{10}
\expandafter\ifx\csname url\endcsname\relax
  \def\url#1{\texttt{#1}}\fi
\expandafter\ifx\csname urlprefix\endcsname\relax\def\urlprefix{URL }\fi
\expandafter\ifx\csname href\endcsname\relax
  \def\href#1#2{#2} \def\path#1{#1}\fi

\bibitem{Lutz:2008km}
M.~F.~M. Lutz, S.~Leupold, {On the radiative decays of light vector and
  axial-vector mesons}, Nucl. Phys. A813 (2008) 96--170.

\bibitem{Terschlusen:2012xw}
C.~Terschl{\"u}sen, S.~Leupold, M.~F.~M. Lutz, {Electromagnetic Transitions in
  an Effective Chiral Lagrangian with the $\eta'$ and Light Vector Mesons},
  Eur. Phys. J. A48 (2012) 190.
\newblock \href {http://dx.doi.org/10.1140/epja/i2012-12190-6}
  {\path{doi:10.1140/epja/i2012-12190-6}}.

\bibitem{Bavontaweepanya:2018yds}
R.~Bavontaweepanya, X.-Y. Guo, M.~F.~M. Lutz, {On the chiral expansion of
  vector meson masses}, Phys. Rev. D98~(5) (2018) 056005.
\newblock \href {http://arxiv.org/abs/1801.10522} {\path{arXiv:1801.10522}},
  \href {http://dx.doi.org/10.1103/PhysRevD.98.056005}
  {\path{doi:10.1103/PhysRevD.98.056005}}.

\bibitem{Lutz:2018cqo}
M.~F.~M. Lutz, Y.~Heo, X.-Y. Guo, {On the convergence of the chiral expansion
  for the baryon ground-state masses}, Nucl. Phys. A977 (2018) 146--207.
\newblock \href {http://arxiv.org/abs/1801.06417} {\path{arXiv:1801.06417}},
  \href {http://dx.doi.org/10.1016/j.nuclphysa.2018.05.007}
  {\path{doi:10.1016/j.nuclphysa.2018.05.007}}.

\bibitem{Guo:2018kno}
X.-Y. Guo, Y.~Heo, M.~F.~M. Lutz, {On chiral extrapolations of charmed meson
  masses and coupled-channel reaction dynamics}, Phys. Rev. D98~(1) (2018)
  014510.
\newblock \href {http://arxiv.org/abs/1801.10122} {\path{arXiv:1801.10122}},
  \href {http://dx.doi.org/10.1103/PhysRevD.98.014510}
  {\path{doi:10.1103/PhysRevD.98.014510}}.

\bibitem{Jenkins:1995vb}
E.~E. Jenkins, A.~V. Manohar, M.~B. Wise, {Chiral perturbation theory for
  vector mesons}, Phys. Rev. Lett. 75 (1995) 2272--2275.
\newblock \href {http://arxiv.org/abs/hep-ph/9506356}
  {\path{arXiv:hep-ph/9506356}}, \href
  {http://dx.doi.org/10.1103/PhysRevLett.75.2272}
  {\path{doi:10.1103/PhysRevLett.75.2272}}.

\bibitem{Klingl:1996by}
F.~Klingl, N.~Kaiser, W.~Weise, {Effective Lagrangian approach to vector
  mesons, their structure and decays}, Z. Phys. A356 (1996) 193--206.
\newblock \href {http://arxiv.org/abs/hep-ph/9607431}
  {\path{arXiv:hep-ph/9607431}}, \href
  {http://dx.doi.org/10.1007/s002180050167} {\path{doi:10.1007/s002180050167}}.

\bibitem{Birse:1996hd}
M.~C. Birse, {Effective chiral Lagrangians for spin 1 mesons}, Z. Phys. A355
  (1996) 231--246.
\newblock \href {http://arxiv.org/abs/hep-ph/9603251}
  {\path{arXiv:hep-ph/9603251}}, \href
  {http://dx.doi.org/10.1007/s002180050105} {\path{doi:10.1007/s002180050105}}.

\bibitem{Bijnens:1997ni}
J.~Bijnens, P.~Gosdzinsky, P.~Talavera, {Vector meson masses in chiral
  perturbation theory}, Nucl. Phys. B501 (1997) 495--517.
\newblock \href {http://arxiv.org/abs/hep-ph/9704212}
  {\path{arXiv:hep-ph/9704212}}, \href
  {http://dx.doi.org/10.1016/S0550-3213(97)00391-X}
  {\path{doi:10.1016/S0550-3213(97)00391-X}}.

\bibitem{Bijnens:1997rv}
J.~Bijnens, P.~Gosdzinsky, P.~Talavera, {Matching the heavy vector meson
  theory}, JHEP 01 (1998) 014.
\newblock \href {http://arxiv.org/abs/hep-ph/9708232}
  {\path{arXiv:hep-ph/9708232}}, \href
  {http://dx.doi.org/10.1088/1126-6708/1998/01/014}
  {\path{doi:10.1088/1126-6708/1998/01/014}}.

\bibitem{Cirigliano:2003yq}
V.~Cirigliano, G.~Ecker, H.~Neufeld, A.~Pich, {Meson resonances, large N(c) and
  chiral symmetry}, JHEP 06 (2003) 012.
\newblock \href {http://arxiv.org/abs/hep-ph/0305311}
  {\path{arXiv:hep-ph/0305311}}, \href
  {http://dx.doi.org/10.1088/1126-6708/2003/06/012}
  {\path{doi:10.1088/1126-6708/2003/06/012}}.

\bibitem{Rosell:2004mn}
I.~Rosell, J.~J. Sanz-Cillero, A.~Pich, {Quantum loops in the resonance chiral
  theory: The Vector form-factor}, JHEP 08 (2004) 042.
\newblock \href {http://arxiv.org/abs/hep-ph/0407240}
  {\path{arXiv:hep-ph/0407240}}, \href
  {http://dx.doi.org/10.1088/1126-6708/2004/08/042}
  {\path{doi:10.1088/1126-6708/2004/08/042}}.

\bibitem{Bruns:2013tja}
P.~C. Bruns, L.~Greil, A.~Sch{\"a}fer, {Chiral behavior of vector meson self
  energies}, Phys. Rev. D88 (2013) 114503.
\newblock \href {http://dx.doi.org/10.1103/PhysRevD.88.114503}
  {\path{doi:10.1103/PhysRevD.88.114503}}.

\bibitem{PACS-CS}
S.~Aoki, et~al., {2+1 Flavor Lattice QCD toward the Physical Point}, Phys. Rev.
  D79 (2009) 034503.
\newblock \href {http://arxiv.org/abs/0807.1661} {\path{arXiv:0807.1661}},
  \href {http://dx.doi.org/10.1103/PhysRevD.79.034503}
  {\path{doi:10.1103/PhysRevD.79.034503}}.

\bibitem{QCDSF-UK}
W.~Bietenholz, et~al., {Flavour blindness and patterns of flavour symmetry
  breaking in lattice simulations of up, down and strange quarks}, Phys. Rev.
  D84 (2011) 054509.
\newblock \href {http://arxiv.org/abs/1102.5300} {\path{arXiv:1102.5300}},
  \href {http://dx.doi.org/10.1103/PhysRevD.84.054509}
  {\path{doi:10.1103/PhysRevD.84.054509}}.

\bibitem{HSC}
H.-W. Lin, et~al., {First results from 2+1 dynamical quark flavors on an
  anisotropic lattice: Light-hadron spectroscopy and setting the strange-quark
  mass}, Phys. Rev. D79 (2009) 034502.
\newblock \href {http://arxiv.org/abs/0810.3588} {\path{arXiv:0810.3588}},
  \href {http://dx.doi.org/10.1103/PhysRevD.79.034502}
  {\path{doi:10.1103/PhysRevD.79.034502}}.

\bibitem{Dudek:2013yja}
J.~J. Dudek, R.~G. Edwards, P.~Guo, C.~E. Thomas, {Toward the excited isoscalar
  meson spectrum from lattice QCD}, Phys. Rev. D88~(9) (2013) 094505.
\newblock \href {http://arxiv.org/abs/1309.2608} {\path{arXiv:1309.2608}},
  \href {http://dx.doi.org/10.1103/PhysRevD.88.094505}
  {\path{doi:10.1103/PhysRevD.88.094505}}.

\bibitem{Bulava:2016mks}
J.~Bulava, B.~Fahy, B.~Hoerz, K.~J. Juge, C.~Morningstar, C.~H. Wong, {$I=1$
  and $I=2$ $\pi-\pi$ scattering phase shifts from $N_{\mathrm{f}} = 2+1$
  lattice QCD}, Nucl. Phys. B910 (2016) 842--867.
\newblock \href {http://arxiv.org/abs/1604.05593} {\path{arXiv:1604.05593}},
  \href {http://dx.doi.org/10.1016/j.nuclphysb.2016.07.024}
  {\path{doi:10.1016/j.nuclphysb.2016.07.024}}.

\bibitem{Brett:2018jqw}
R.~Brett, J.~Bulava, J.~Fallica, A.~Hanlon, B.~Hoerz, C.~Morningstar,
  {Determination of $s$- and $p$-wave $I=1/2$ $K\pi$ scattering amplitudes in
  $N_{\mathrm{f}}=2+1$ lattice QCD}, Nucl. Phys. B932 (2018) 29--51.
\newblock \href {http://arxiv.org/abs/1802.03100} {\path{arXiv:1802.03100}},
  \href {http://dx.doi.org/10.1016/j.nuclphysb.2018.05.008}
  {\path{doi:10.1016/j.nuclphysb.2018.05.008}}.

\bibitem{LHPC}
A.~Walker-Loud, et~al., {Light hadron spectroscopy using domain wall valence
  quarks on an Asqtad sea}, Phys. Rev. D79 (2009) 054502.
\newblock \href {http://arxiv.org/abs/0806.4549} {\path{arXiv:0806.4549}},
  \href {http://dx.doi.org/10.1103/PhysRevD.79.054502}
  {\path{doi:10.1103/PhysRevD.79.054502}}.

\bibitem{Lutz:2014oxa}
M.~F.~M. Lutz, R.~Bavontaweepanya, C.~Kobdaj, K.~Schwarz, {Finite volume
  effects in the chiral extrapolation of baryon masses}, Phys. Rev. D90~(5)
  (2014) 054505.
\newblock \href {http://dx.doi.org/10.1103/PhysRevD.90.054505}
  {\path{doi:10.1103/PhysRevD.90.054505}}.

\bibitem{Patrignani:2016xqp}
C.~Patrignani, et~al., {Review of Particle Physics}, Chin. Phys. C40~(10)
  (2016) 100001.
\newblock \href {http://dx.doi.org/10.1088/1674-1137/40/10/100001}
  {\path{doi:10.1088/1674-1137/40/10/100001}}.

\bibitem{Geneva}
R.~Berlich, S.~Gabriel, A.~Garcia, M.~Kunze, \href{www.gemfony.eu}{{Distributed
  Parametric Optimization with the Geneva Library}}, Data Driven e-Science,
  Conference proceedings of ISGC 2010, Springer New York (2010) 303.
\newline\urlprefix\url{www.gemfony.eu}

\bibitem{Bijnens:2014lea}
J.~Bijnens, G.~Ecker, {Mesonic low-energy constants}, Ann. Rev. Nucl. Part.
  Sci. 64 (2014) 149--174.
\newblock \href {http://arxiv.org/abs/1405.6488} {\path{arXiv:1405.6488}},
  \href {http://dx.doi.org/10.1146/annurev-nucl-102313-025528}
  {\path{doi:10.1146/annurev-nucl-102313-025528}}.

\bibitem{Bornyakov:2016dzn}
V.~G. Bornyakov, R.~Horsley, Y.~Nakamura, H.~Perlt, D.~Pleiter, P.~E.~L. Rakow,
  G.~Schierholz, A.~Schiller, H.~Stüben, J.~M. Zanotti, {Flavour breaking
  effects in the pseudoscalar meson decay constants}, Phys. Lett. B767 (2017)
  366--373.
\newblock \href {http://arxiv.org/abs/1612.04798} {\path{arXiv:1612.04798}},
  \href {http://dx.doi.org/10.1016/j.physletb.2017.02.018}
  {\path{doi:10.1016/j.physletb.2017.02.018}}.

\bibitem{Dowdall:2013rya}
R.~J. Dowdall, C.~T.~H. Davies, G.~P. Lepage, C.~McNeile, {Vus from pi and K
  decay constants in full lattice QCD with physical u, d, s and c quarks},
  Phys. Rev. D88 (2013) 074504.
\newblock \href {http://arxiv.org/abs/1303.1670} {\path{arXiv:1303.1670}},
  \href {http://dx.doi.org/10.1103/PhysRevD.88.074504}
  {\path{doi:10.1103/PhysRevD.88.074504}}.

\bibitem{Bruno:2016plf}
M.~Bruno, T.~Korzec, S.~Schaefer, {Setting the scale for the CLS $2 + 1$ flavor
  ensembles}, Phys. Rev. D95~(7) (2017) 074504.
\newblock \href {http://arxiv.org/abs/1608.08900} {\path{arXiv:1608.08900}},
  \href {http://dx.doi.org/10.1103/PhysRevD.95.074504}
  {\path{doi:10.1103/PhysRevD.95.074504}}.

\bibitem{Guo:2018c}
X.-Y. Guo, M.~F.~M. Lutz, {On the chiral SU(3) extrapolations of the pion and
  kaon decay constants. }\href {http://arxiv.org/abs/1810.07376}
  {\path{arXiv:1810.07376}}.

\bibitem{Andersen:2018mau}
C.~Andersen, J.~Bulava, B.~Hörz, C.~Morningstar, {The $I=1$ pion-pion
  scattering amplitude and timelike pion form factor from $N_{\rm f} = 2+1$
  lattice QCD. }\href {http://arxiv.org/abs/1808.05007}
  {\path{arXiv:1808.05007}}.

\end{thebibliography}

\end{document}